\documentclass[a4paper]{jpconf}
\usepackage{graphicx}
\usepackage[latin1]{inputenc}

\begin{document}

\title{Solution-space structure of (some) optimization problems}

\author{Alexander K. Hartmann$^1$, Alexander Mann$^2$ and Wolfgang
  Radenbach$^3$}

\address{$^1$ Institute for Physics, University of Oldenburg, 26111
  Oldenburg, Germany}

\address{$^2$ Institute for Theoretical Physics, University of
  G\"ottingen, Friedrich-Hund-Platz 1, 37077 G\"ottingen, Germany}

\address{$^3$ University of G\"ottingen, Platz der G\"ottinger Sieben
  5, 37073 G\"ottingen, Germany}

\begin{abstract}
We study numerically the cluster structure of random ensembles of  two
NP-hard optimization problems originating in computational complexity,
the vertex-cover problem and the number partitioning problem. We use
branch-and-bound type algorithms to obtain exact solutions of these problems
for moderate system sizes. Using two methods, 
direct neighborhood-based clustering and
hierarchical clustering, we investigate the structure of the solution
space. The main result is that the correspondence between solution structure
and the phase diagrams of the problems is not unique. Namely, for vertex
cover
we observe a drastic change of the solution space from large single
clusters to multiple nested levels of clusters. In contrast, for the
number-partitioning problem, the phase space looks always very
simple, similar to a random distribution of the lowest-energy
configurations. This
holds in the ``easy''/solvable phase as well as in the
``hard''/unsolvable phase. 
\end{abstract}

\section{Introduction}

For NP-hard optimization problems \cite{garey1979} no algorithm is
known which solves the problem in the worst case in a time which 
grows less than exponentially with the problem size. From a physical point
of view, this implies the notion that the structure of the underlying
solution space is probably organized in a complicated way, e.g.\
characterized by a hierarchical clustering. Such clustering has already been
observed in statistical physics when studying spin glasses 
\cite{binder1986,mezard1987,fischer1991,young1998}. 
For the mean-field Ising spin glass, also
called Sherrington-Kirkpatrick (SK) model \cite{sherrington1975}, the 
solution
exhibits replica-symmetry breaking
(RSB) \cite{parisi1979,parisi1983}, 
which means that the state space is organized in an infinitely
nested hierarchy of clusters of states, characterized by
ultrametricity \cite{rammal1986}.
Also in numerical
studies the clustering structure of the SK model has been observed,
e.g. by calculating the distribution of overlaps
\cite{young1983,parisi1993,billoire2003},  when studying
the spectrum of spin-spin correlation matrices
\cite{sinova2000,sinova2001} or when applying direct clustering
\cite{hed2004}. On the other hand,
for models like Ising ferromagnets it is clear that they 
do not exhibit RSB and all solutions are organized in one cluster 
(resp. two, when including spin-flip symmetry).

The use of notions and analytical tools 
from statistical mechanics enabled physicists recently to contribute
to the analysis of these NP-hard problems that
 originate in theoretical computer
science.  Well known problems of this kind are the satisfiability (SAT)
problem \cite{biroli2000, mezard2002, mezard2002b}, 
number partitioning  \cite{mertens1998,mertens2000},
graph coloring \cite{mulet2002}, and vertex
cover \cite{cover2000,cover-tcs2001,cover-long2001,zhou2003,cover-review2003}. 
Analytically, the phase diagrams of these
problems on suitably defined parametrized ensembles of random
instances can be studied using some well-known techniques from
statistical physics, like the replica trick \cite{monasson1996,monasson1999}, 
or  the cavity approach
\cite{mezard2002b}. But full solutions have not been found 
in most cases,
since these problems are usually not
defined on complete graphs but on diluted graphs, 
which poses additional technical
problems. Usually, one can only calculate the
solution in the case of replica symmetry 
\cite{monasson1996,monasson1999,biroli2000},
or in the case of one-step
replica symmetry breaking (1-RSB) \cite{mezard2002,mezard2002b}, 
and look for the stability of the solutions.
For this reason, the relation between the
solution and the clustering structure is not well established and it
is far from being clear for most models 
how the clustering structure looks like.
However, most statistical physicist believe that the failure of
replica symmetry (RS)
leads indeed to clustering \cite{weigt2004,zecchina2004}.
So far, only few analytical 
studies of the clustering properties of classical combinatorial 
optimization problems like SAT have been performed
\cite{biroli2000,mezard2003}. 
These results depend or may depend on the specific
assumptions one makes when applying certain analytical tools and when
performing approximations.
In particular, it is unlikely that the
clustering of models on dilute graphs is exactly the same as it is
found for the mean-field SK spin glass.
 So from the
physicist's point of view, it is quite interesting to study the
organization of the phase space using numerical methods 
to understand better the
meaning of ``complex organization of phase space'' for other,
non-mean-field models, like combinatorial optimization problems.
It is
the aim of this paper to study numerically 
the clustering properties of two
particular problems, the vertex-cover problem and the number partitioning
problem.

The study of the solution structure is not only important for
physicists, but also of interest for computer
science. 
From an algorithmic point of view, especially 
the solution-space structure seems to
play an important role. However, as we will see in this work,
a direct one-to-one correspondence between cluster structure
and hardness of the problem cannot be described at the moment.

\section{Models}

In this work, we deal with two classical problems
\cite{garey1979} from
computational complexity, the {\em vertex-cover problem} (VC)
and the {\em number partitioning problem} (NPP).

VC is defined for an arbitrary given
graph $G=(V,E)$, $V$ being a set of $N$ vertices and $E\subset
V\times V$ a set of undirected edges.
Let $V' \subset V$ be a subset of all vertices. We call a vertex $v$
\emph{covered} if $v\in V'$, \emph{uncovered} if $v\notin V'$.
Similarly an edge $\{i,j\}$
 is \emph{covered} if at least one of
its endpoints $i,j$ is covered. 
If all edges of $G$ are covered, then we call $V'$ a vertex cover
$V_{VC}$.  We denote $X\equiv|V'|$ and $x\equiv X/N$.
We describe each subset $V'$ by a configuration 
vector ${\vec{x}}\in\{0,1\}^N$ with
$x_i=1\, \leftrightarrow\, i \in V'$. 
For a graph $G=(V,E)$  the minimal VC problem is
the following optimization problem: Construct a vertex cover
$V_{VC-min}\subset V$ of
minimal cardinality and find its size
$X_{\min}\equiv \left|V_{VC-min}\right|$. 
Usually there are many solutions of the same size, hence the solutions
are {\em degenerate}.

Here we study VC  on an ensemble of random graphs  with
$V= \{1, 2, \dots N\}$ and  $\frac c 2 N$ randomly drawn, undirected edges
$\left\{i,j\right\} \in E$.
 In this notation $c$ is the connectivity, i.\,e. the
average number of edges each vertex is contained in. For this
ensemble,
one can calculate the average fraction $X_{\min}/N$ of vertices that
have to be covered for minimum vertex covers by using methods
from statistical mechanics \cite{cover2000} like the replica approach 
\cite{mezard1987,fischer1991}, the cavity approach \cite{zhou2003}
or by an analysis of a heuristic algorithm, called leaf-removal
\cite{bauer2001},  which approximately  solves the problem
\cite{bauer2001a}. The main result is that for $c\le e \approx 2.718$, 
$x_{\min}$ can be obtained exactly, while for $c>e$ only approximate
solutions have been found so far. Within the statistical mechanics treatment 
\cite{cover2000}, this means that the solutions for $c\le e$ 
are RS,
while for larger connectivities RSB
appears.
It has been found \cite{vccluster2004} that the onset of RSB can be
seen numerically from analyzing the cluster structure of the solution space.
Note that the leaf-removal heuristic \cite{bauer2001} allows, in
conjunction with an exact branch-and-bound approach \cite{opt-phys2001}, to
solve VC on random graphs for $c\le e$ typically in polynomial
time. For $c>e$, one needs typically an exponential running
time. Hence, in this case, the onset of a complex solution landscape,
visible analytically \cite{cover2000} and numerically as presented in this
work,  coincides with an ``easy-hard transition'' of the typical
computational complexity.

The NPP is defined for a given set ${\cal A}$ of $N$ natural positive $M$-bit
numbers $0\le a_i \le 2^M-1$. The NPP is the optimization problem 
to find a partition of
${\cal A}$ into two subsets ${\cal A'}$ and ${\cal A\setminus A'}$
such that the difference (or {\em energy}) $E=|\sum_{a_i\in{\cal A'}} a_i -
\sum_{a_i\in{\cal A\setminus A'}} a_i|$ is minimal. 
Similar to VC, 
a partition can also be described by a vector ${\vec{x}}\in\{0,1\}^N$ with,
in this case,
$x_i=1\, \leftrightarrow\, a_i \in {\cal A'}$. Note that partitions
with $E=0$ are called {\em perfect}.

Here we study NPP on an ensemble of random numbers, which are distributed
equally in $[0,2^M-1]$. It has been observed numerically \cite{gent1996}, 
within
a statistical mechanics treatment \cite{mertens1998}
and also using an exact mathematical analysis \cite{borgs2001}, 
that for $\kappa\equiv
M/N$ larger than a critical value $\kappa_c=\kappa_c(N)$ ($\kappa_c\to
1$ for $N\to \infty$) almost surely no perfect partitions exist, while
for $\kappa <\kappa_c(N)$ there are typically exponentially many perfect 
partitions. This transition coincides with an ``easy-hard transition''
of an exact algorithm \cite{korf1998}, i.e.\ for a fixed value 
$\kappa<\kappa_c$
perfect partitions can be found typically in a time polynomially 
increasing with $N$ , while
for $\kappa>\kappa_c$, it takes an exponentially growing running time
to find the minimum partition. Below, we will analyze the cluster
structure of NPP. To study the behavior as a function of system size, since
$\kappa_c$ depends somehow on $N$, we use $k=\kappa/\kappa_c(N)$ as parameter
indicating the distance from the phase boundary. Our main result is that
the behavior does {\em not} differ significantly
below and above $k=1$, in contrast to the results for VC.

\section{Algorithms}

\label{sec:algos}

We apply exact algorithms, which guarantee to find the optimum solution.
These algorithms are based on the branch-and-bound approach
\cite{lawler66}. The basic idea is, as each variable $x_i$ of a 
configuration is either 0 or 1, there are $2^N$ possible
configurations which can be arranged as leaves of a binary
({\em branching}) tree. At each node of the tree, the two subtrees
represent the subproblems where the corresponding variable is either
0 or 1.  The algorithm constructs this tree partially, while searching for
a solution.
A subtree will be omitted if its leaves can be
proven to contain less favorable configurations than the best of all previously
considered configurations, this is the so-called {\em bound}. The actual
performance of a branch-and-bound algorithm depends strongly on
the heuristic which is used to decide in which order the variables
are assigned and on the bounds used. Both have to be chosen 
according to the problem under consideration.

For VC, we use the branch-and-bound algorithm presented in references 
\cite{vccluster2004,opt-phys2001,phase-transitions2005}, which is based
on reference
 \cite{tarjan1977}. For NPP, we apply the Korf algorithm \cite{korf1998}
with a bound described in reference \cite{mertens2006}.

\section{Clustering Algorithms}

We apply two methods to study the cluster structure of the solution landscape,
neighborhood-based clustering and hierarchical clustering.

The neighborhood-based  
approach   is
based on the hamming distance between different solutions. The hamming
distance
$dist_{ham}({\vec{x}}^{(\alpha)},{\vec{x}}^{(\beta)}) \equiv
d_{\alpha\beta}$ of two solutions
 is the number of variables in which the two configurations differ, i.e.\
$d_{\alpha\beta} = \sum_i | x^{(\alpha)}_i-x^{(\beta)}_i|$. If for
 two optimal solutions their hamming distance is not larger  than a given
 value $d_{\max}$,
  we will call them \emph{neighbors}.  For VC, we use $d_{\max}=2$ (the minimal possible distance of nonidentical configurations),
  while for the NPP, $d_{\max}$ will depend on the system size $N$,
  see below.

We define a \emph{cluster} $\cal C$ as maximal set of solutions, that can
be reached by repeatedly moving to neighboring solutions.
States which
belong to different clusters are separated by a hamming distance of at
least $d_{\max}+2$ for VC resp. $d_{\max}+1$ for NPP. 
Similar definitions of clusters have been used e.g. for the 
analysis of random p-XOR-SAT \cite{mezard2003} or finite-dimensional
spin glasses \cite{valleys-long}.

To decide whether two arbitrary solutions
 ${\vec{x}}^{(\alpha)}$ and ${\vec{x}}^{(\beta)}$
belong to the same cluster or not, one needs to calculate the complete
cluster ${\vec{x}}^{(\alpha)}$ (or ${\vec{x}}^{(\beta)}$) belongs
to. Hence the clustering is very expensive. 

The neighborhood-based algorithm is as follows, given a complete set
$S$ of all solutions. 
\begin{tabbing}
xxx\=xxx\=xxx\=xxx\=xxx\=xxx\=xxx\=xxx\=\kill
{\bf begin}\\
\>$i = 0$ \{number of so far detected clusters\}\\
\>{\bf while} $S$ not empty {\bf do}\\
\> {\bf begin}\\
\>\>$i = i + 1$\\
\>\>remove an element ${\vec{x}}^{(\alpha)}$ from $S$\\
\>\>set cluster $C_i=({\vec{x}}^{(\alpha)})$\\
\>\>set pointer ${\vec{x}}^{(\beta)}$ to first element of $C_i$\\
\>\>{\bf while ${\vec{x}}^{(\beta)} <> NULL$} {\bf do}\\
\>\> {\bf begin}\\
\>\>\>{\bf for all} elements ${\vec{x}}^{(\gamma)}$ of $S$\\
\>\>\>\>{\bf if} $d_{ham}({\vec{x}}^{(\beta)},
{\vec{x}}^{(\gamma)})\le d_{\max}$ 
{\bf then}\\
\>\>\>\>{\bf begin}\\
\>\>\>\>\>remove ${\vec{x}}^{(\gamma)}$ from $S$\\
\>\>\>\>\>put ${\vec{x}}^{(\gamma)}$ at the end of $C_i$\\
\>\>\>\>{\bf end}\\
\>\>\>set pointer ${\vec{x}}^{(\beta)}$ to next element of $C_i$\\
\>\>\> or to $NULL$ if there is no more\\
\>\>{\bf end}\\
\>{\bf end}\\
{\bf end}
\end{tabbing}

 The
crucial point is that one really needs to consider all solutions
and not just a sample. The algorithm is quadratic in the number of
solutions $\{\vec{x}^{(\alpha)}\}$,  which makes the method
applicable to system 
sizes, depending on the actual problem, of the order of $N \approx 100$.

As an alternative method, we will use a clustering approach that
organizes the states in a hierarchical structure. Such clustering
methods \cite{jain1988} are widely used in general data analysis,
sometimes also used in statistical mechanics, see e.g. references 
\cite{hed2001,ciliberti2003,hed2004}. The methods all 
start by assuming that all states belong to
separate clusters. Similarity between clusters (and states) is defined
by a measure called \emph{proximity matrix} $\tilde d_{\alpha,\beta}$.  At
each step two very similar clusters are joined and so a hierarchical tree of
clusters is formed.
As proximity measure for two initial clusters, each
containing only a single state, we naturally choose the hamming
distance between these two states as defined above, divided by the number 
of vertices: $\tilde d_{\alpha,\beta}=d_{\alpha\beta}/N$. 
At each step the two clusters $C_\alpha$
and $C_\beta$ with the minimal distance are merged to form a new cluster
$C_\gamma$. Then the proximity matrix is updated by deleting the
distances involving $C_\alpha$ and $C_\beta$ and adding the
distances between $C_\gamma$ and all other clusters $C_\delta$ in the
system. So we need to extend the proximity measure to clusters
with more than one state, based on some suitable update rule which is
usually a function of the 
distances $\tilde d_{\alpha,\beta}$, $\tilde d_{\alpha,\delta}$ and 
$\tilde d_{\beta,\delta}$.

The choice of this function is a widely discussed field since it can
have a great impact on the clustering obtained \cite{jain1988}. It should
represent the natural organization present in the data and 
not some artificial structure induced from the choice of the update
rule. 
Here we will use \emph{Ward's method} (also called 
\emph{minimum-variance method}) \cite{ward1963}. 
The distance between the merged cluster
$C_\gamma$ and some other cluster $C_\delta$ is given by
\begin{equation}
  \label{eq:warddist}
  \tilde d_{\gamma,\delta} =
  \frac{(n_\alpha+n_\delta)\tilde d_{\alpha,\delta}+
(n_\beta+n_\delta)\tilde d_{\beta,\delta} -
n_\delta \tilde d_{\alpha,\beta} }{n_\alpha+n_\beta+n_\delta}\,,
\end{equation}
where $n_\alpha, n_\beta, n_\delta$ are the number of elements in
cluster $C_\alpha,C_\beta,C_\delta$, respectively.
Heuristically Ward's method seems to outperform other update
rules. The choice guarantees that at each step the two clusters to be
merged are chosen in a way that the variance inside each cluster summed
over all clusters increases by the minimal possible amount. 

The output of the clustering algorithm can be represented as  a
\emph{dendrogram}. This is a tree with the configurations as leaves and % dendRogram laut Wikipedia und Google (alle 6 Vorkommen geändert, nur hier markiert)
each node representing one of the clusters at different levels of
hierarchy, see the bottom half of the examples in figure \ref{fig:dendo}.
Note that Ward's algorithm is able to cluster any data. Even if no structure were present, the data could always be displayed as a dendrogram.
Hence, one has to perform additional checks.  Here, we use a visual
check, i.e.\ we
plot the hamming distances as a matrix where the rows and columns are
ordered according to the dendrogram.

\section{Results}

We first summarize 
our results \cite{vccluster2004} 
for the solution-space analysis for VC. Second, we present the data
obtained for the NPP.

\subsection{Vertex cover}

\begin{figure}[htbp]
  \centering
\includegraphics[width = 0.6\textwidth,clip]{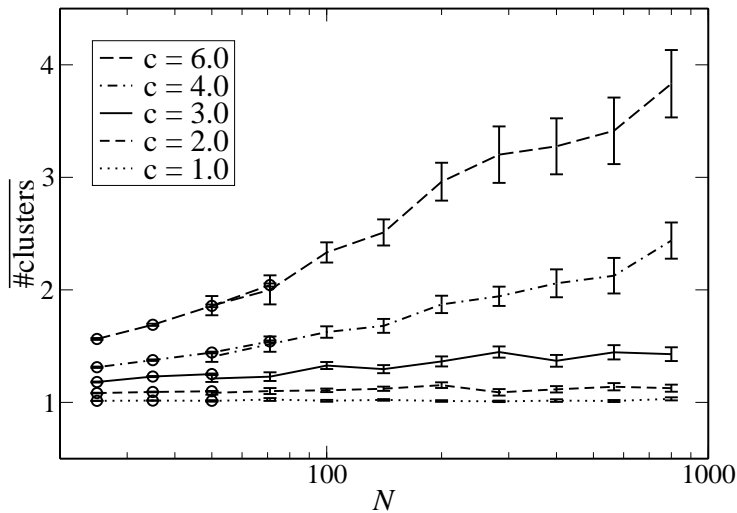}  
  \caption{Average number of clusters in the solution space of the largest
    connected component as function of system size. The
    circle symbols for small system sizes have been obtained by
    clustering complete sets of solutions. For large systems we
    sampled solutions with a Monte Carlo algorithm at large but
    finite chemical potential $\mu$, for details see reference
 \cite{vccluster2004}.}
  \label{fig:clusterneighbour}
\end{figure}

For every value of $N$ we sampled $10^4$ realizations. 
For each realization, we considered only the largest connected
component of the given graph, since the solution-space structures of 
different connected components are independent of each other.
The average
number of clusters, as obtained from the neighborhood-based
clustering using $d_{\max} =2$, as a function of the connectivity is
shown as circles in figure
\ref{fig:clusterneighbour}. For larger system sizes, the number of cluster
was estimated as described in
reference \cite{vccluster2004}.

 For $c<e$ the number of clusters remains
close to one. For larger values of $c$ the number of clusters increases
with system size. Apparently the increase is compatible with a
logarithmic growth as a function of system size. Hence, the change
from simple to complex behavior, as expected from the analytical
results \cite{cover2000} is visible through the cluster analysis.

\begin{figure*}[t]
  \centering
\begin{minipage}[t]{0.49\textwidth}
\centering
\vspace*{-3mm}

 \includegraphics[width =0.8\textwidth]{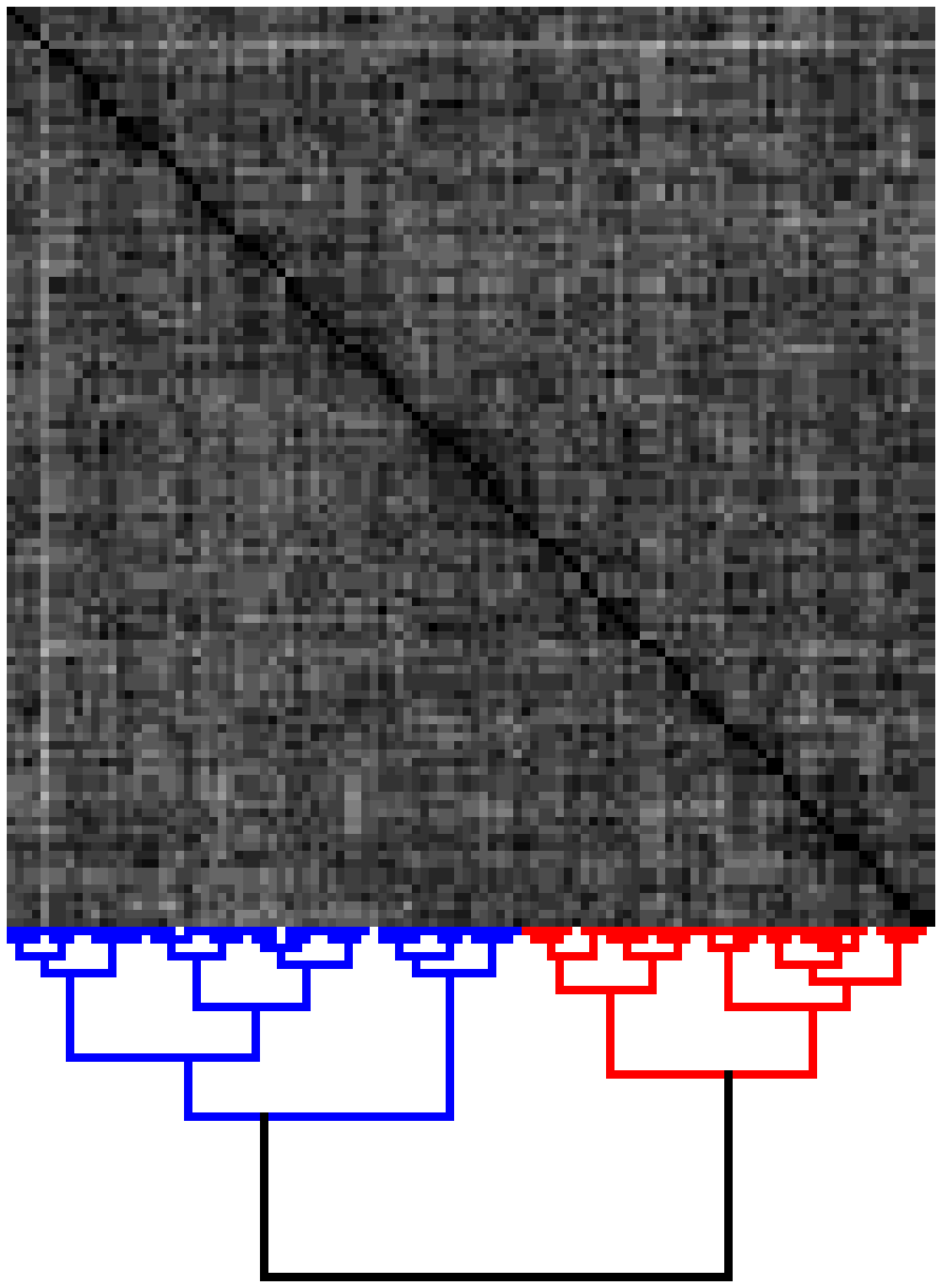} 
 \end{minipage}
 \begin{minipage}[t]{0.49\textwidth}
\centering
\vspace*{-3mm}

   \includegraphics[width =0.8\textwidth]{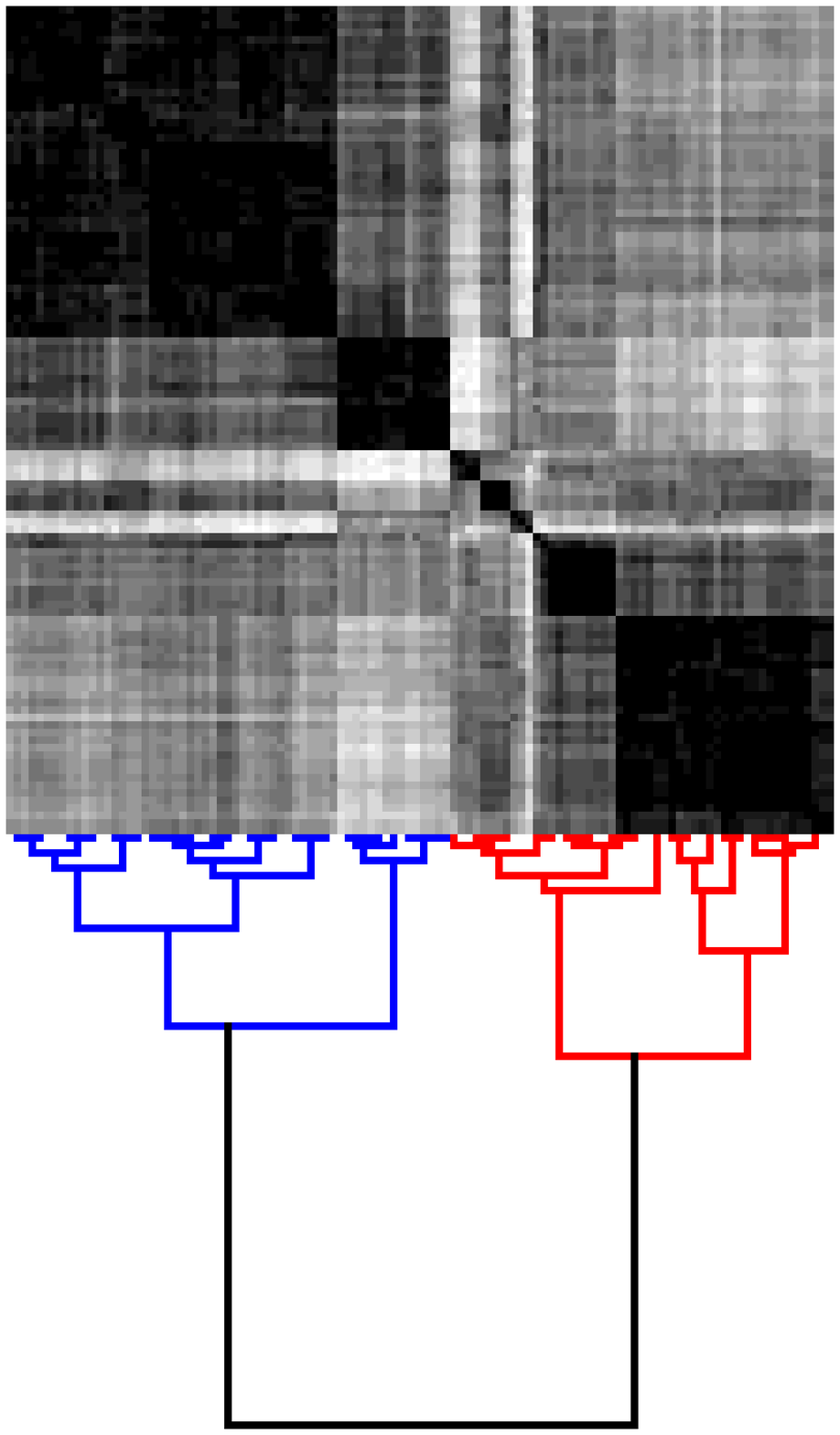}  
 \end{minipage}
  \caption{Sample dendrograms of 100 VC solutions for a graph with 400
    vertices. Darker colors correspond to closer distances. The left
    one is at $c=2$, i.\,e. in the RS phase. There is no
    structure present.  For
    $c=6$ the dendrogram provides a structure, where the
    solutions form clusters. The careful reader may recognize a
    second or third level of clustering in the right picture.}
  \label{fig:dendo}
\end{figure*}

The results of the hierarchical clustering for VC are shown in
figure \ref{fig:dendo}. Darker colors correspond to smaller distances.
The figure shows two different realizations: For small values of
$c<e$,
 the system is in the RS phase, only a
single cluster is present. For larger values of $c$, 
the ordering of the states obtained by the clustering algorithm
reveals an underlying structure which can be seen in the right part of
the figure. One can see that the states form groups where the hamming
distance between the members is small (dark colors) while the distance
to other states is large. Thus, our results are compatible with 
clustering being present for realizations
with $c>e$. If you look carefully you can see more structure inside
the clusters. Multiple levels of clustering indicate higher levels of
RSB which we expect to be present for these values of $c$
\cite{cover-long2001,zhou2003}.

\subsection{Number-partitioning problem}

\begin{figure}[htbp]
  \centering
\includegraphics[angle=270,width = 0.6\textwidth,clip]{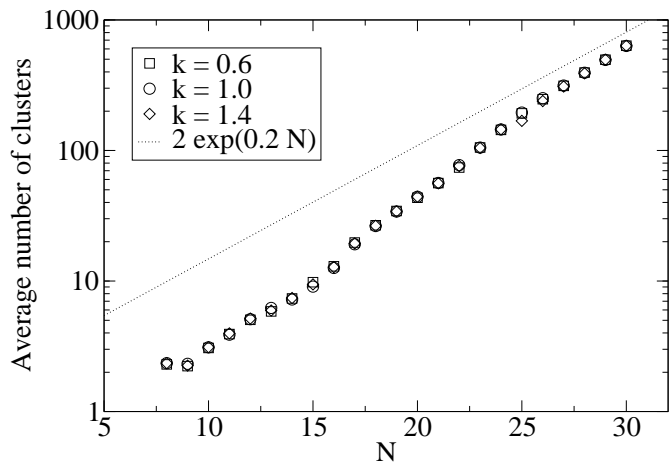}   
% \change{neue Grafik Nr. 3b verwenden!}
  \caption{Average number of clusters in the solution space of the
  $K=2\exp(0.2N)$  energetically lowest-lying configurations 
  as a function of the system size, for three different values of $k$.
Note that $d_{\max}=\sqrt{N}$ has been used.}
  \label{fig:clusterNPP}
\end{figure}

Next, we consider the NPP, to see whether similar clustering and
coincidences can be found there as well. For all values of $k=\kappa/\kappa_c$ 
and all sizes $N$, the $Z=2\exp(0.2N)$ energetically lowest lying
configurations were selected. In the solvable 
phase $k<1$, these are 
perfect partitions, while for $k>1$ we take the $Z$ configurations with the lowest
values of $E$.
In figure \ref{fig:clusterNPP},
the average number of clusters obtained using the neighbor-based clustering 
with $d_{\max}=\sqrt{N}$ is shown. In all cases, in the ``easy'' as well as
in the  ``hard'' phase, a basically exponential increase of the number
of clusters is obtained. 
Note that if instead $d_{\max}=2$ had been chosen, like
in the VC case, an even larger number of clusters would result, hence the
large number of clusters is {\em not} an artifact of the choice of $d_{\max}$.
Hence, from this result, one is tempted to conclude that the solution
landscape is very complex.

\begin{figure*}[t]
  \centering
\begin{minipage}[t]{0.49\textwidth}
\centering
\vspace*{-3mm}

 \includegraphics[width =0.8\textwidth]{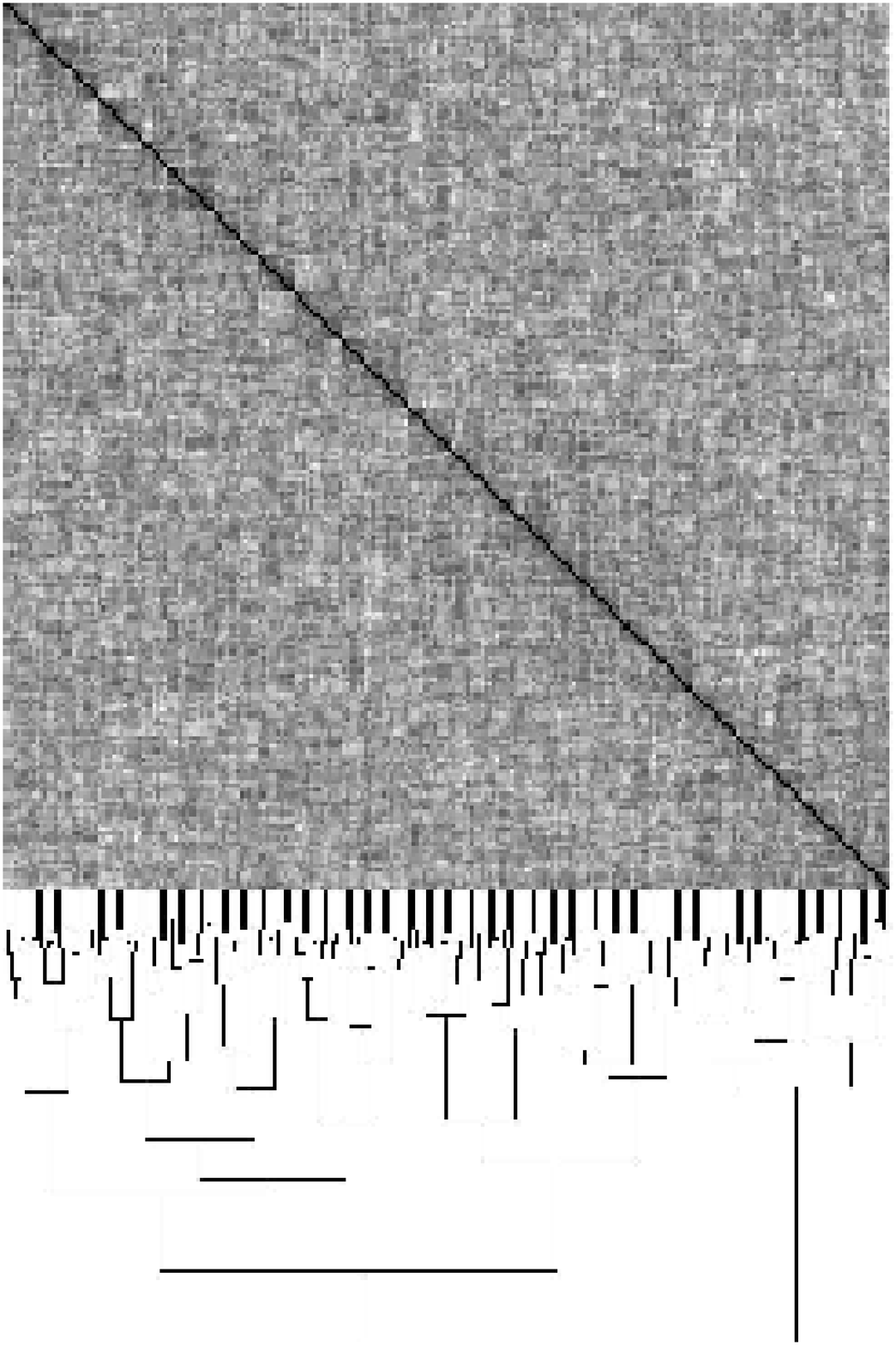} 
 \end{minipage}
 \begin{minipage}[t]{0.49\textwidth}
\centering
\vspace*{-3mm}

   \includegraphics[width =0.8\textwidth]{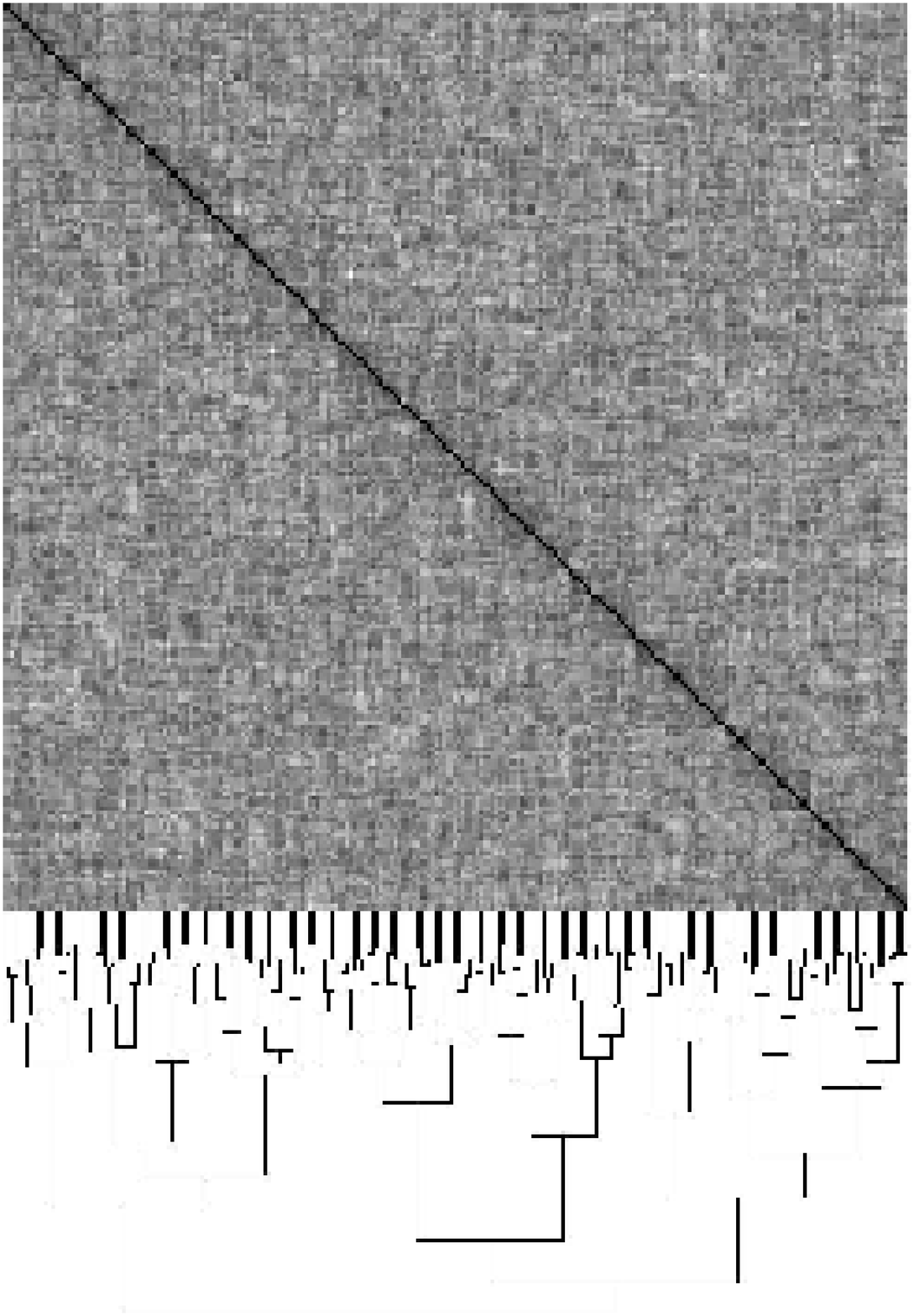}  
 \end{minipage}
  \caption{Sample dendrograms (bottom) and distance matrices (top) 
of 200 NPP solutions for systems with 35
numbers. Darker colors correspond to closer distances. The left
    one is at $k=1.5$, i.\,e. in the hard and unsolvable phase, while
the right is for $k=0.6$, i.e. in the solvable phase, where perfect partitions
exist.  In both cases there is no
    structure present.   }
  \label{fig:dendo2}
\end{figure*}

Nevertheless, when one performs the hierarchical clustering with a 
certain number (200) of randomly selected solutions ($\kappa<\kappa_c$)
or the 200 energetically lowest lying configurations $(\kappa>\kappa_c$),
one obtains a simple uniform distributions of the distances, without any
structure visible, see
figure \ref{fig:dendo2}. This shows, that in both
regions of the phase space,
the solution space basically
consists of equally spaced configurations, which have a large distance 
from each other such that they appear as single-configuration clusters.
Note that we have also studied larger systems
using stochastic algorithms, and again no structure was found.
Hence, the structure of the solution space is in fact simple everywhere,
in contrast to the VC. This might coincide with the notion that the
NPP can be seen \cite{bauke2004} 
as a variant of the random-energy model \cite{derrida1981}, where each 
configuration is assigned a randomly and independently selected energy value.

\begin{figure}[htbp]
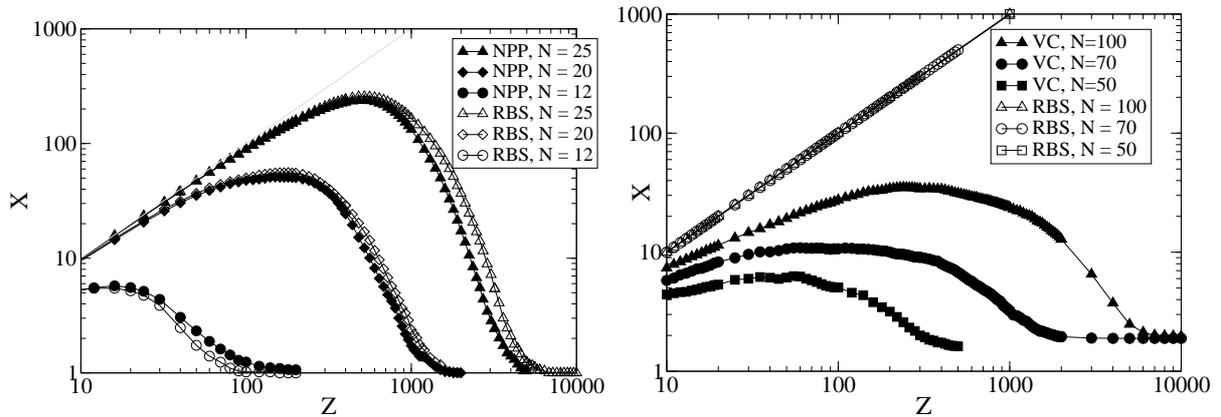

  \centering
\includegraphics[width = 0.49\textwidth,clip]{vergleich_K_bitstr4.eps}
\includegraphics[width = 0.49\textwidth,clip]{vc_rbs.eps}

% \change{neue Grafik Nr. 4 verwenden!}

  \caption{Average number of clusters from neighborhood-based 
clustering (full symbols) 
in the solution space of the
  $Z$ energetically lowest lying configurations 
  as a function of $Z$, for different values of $N$.  
For comparison also the same number of 
clusters is shown, when instead  
$Z$ random bit strings are used (open symbols), 
i.e.\ the assumption of absolutely no order
in the configuration space.
Left: NPP for $k=1.2$, $d_{\max}=\sqrt{N}$. Right: VC 
for $c=4$, $d_{\max}=2$.
}
  \label{fig:comparisonBitStrings}
\end{figure}

To support this finding, we have compared the solution-space structure of the 
NPP to a set of random bit strings (RBS), i.e. to a set of 
configurations with the least possible structure. For this purpose, we again 
use the neighborhood-based clustering with $d_{\max}=\sqrt{N}$ and
measure the average number of clusters (averaged over 100
realizations) as a function of the number $Z$ 
of configurations included in the clustering. We do this for the NPP, for 
$k=1.2$, where the $Z$ energetically lowest lying configurations are considered.
For the RBS case, we use $Z$ independent strings. For the RBS
case, we expect that for small values of $Z$, the average number of clusters
increases and is close to $Z$, because each configuration forms a single
cluster. For large and even more 
increasing values of $Z$, the configurations are lying more dense in
the configuration space, hence the number of configurations which have
a neighbor within the distance $d_{\max}$ increases, leading to a
decrease of the number of clusters. Finally, if $Z$ is large enough,
all configurations are interconnected, hence one obtains just one single 
cluster.
This behavior is exactly visible in figure \ref{fig:comparisonBitStrings}.
Interestingly, the resulting curves for the NPP are following very closely
the curves for the  RBS case, hence one can conclude that the NPP has
a cluster structure which is almost indistinguishable from a random
distribution, supporting the above results. Note that for VC, we find different
results. The number of clusters first increases a bit, and then
decreases to a value clearly above one, hence differs strongly from the RBS
case.

\section{Summary and Outlook}

We have studied the cluster structure of 
two combinatorial optimization problems, the vertex-cover problem and the
number partitioning problem, on suitably defined random ensembles.
In both cases, the existence of phase transitions which coincide with
drastic changes of the computational complexity are  analytically well
established. For many of these problems, exact 
analytical solutions are at least currently out of reach; 
in particular the cluster structure
of the solution landscape has been studied only very partially by analytical 
means. Hence, in this work, we analyze the cluster landscape numerically.
We calculate exact solutions which we study using direct
neighborhood clustering and via a hierarchical clustering approach.
For VC, the ``easy-hard'' phase transition is visible also in the cluster 
structure of the solution space, the hard phase is dominated by a complicated 
hierarchy of solutions. Instead,
for the NPP, the cluster structure is
basically equivalent to a random distribution of the solutions in the 
``easy''/solvable as well as in the ``hard''/unsolvable phase. Hence, a direct
correspondence between the structure of the solution space,
the solvability, and the hardness of finding a (best) solution does not
seem to exist. Hence, the NPP seems to be hard in a different way than VC.
The reason might be that the NPP is a pseudopolynomial problem, i.e. when
fixing the number $M$ of bits, the problems becomes always quickly solvable for
$N\to\infty$ .

To further elucidate possible relationships, other
optimization problems should be studied in this way in the future.
In particular the authors are currently working on the satisfiability
problem, where several changes of the solution-space structure are
expected from approximate analytical solutions \cite{biroli2000,zecchina2004}.

\section{Acknowledgments}

The authors have received financial support from the {\em
  VolkswagenStiftung} (Germany) within the program ``Nachwuchsgruppen
an Universit\"aten'', and from the European Community via the
DYGLAGEMEM program.

\bibliographystyle{iopart-num} 
\bibliography{alex_refs}

\providecommand{\newblock}{}
\begin{thebibliography}{10}
\expandafter\ifx\csname url\endcsname\relax
  \def\url#1{{\tt #1}}\fi
\expandafter\ifx\csname urlprefix\endcsname\relax\def\urlprefix{URL }\fi
\providecommand{\eprint}[2][]{\url{#2}}

\bibitem{garey1979}
Garey M~R and Johnson D~S 1979 {\em Computers and intractability\/} (San
  Francisco: W.H. Freemann)

\bibitem{binder1986}
Binder K and Young A 1986 {\em Rev. Mod. Phys.\/} {\bf 58} 801

\bibitem{mezard1987}
M\'ezard M, Parisi G and Virasoro M 1987 {\em Spin glass theory and beyond\/}
  (Singapore: World Scientific)

\bibitem{fischer1991}
Fischer K and Hertz J 1991 {\em Spin Glasses\/} (Cambridge: Cambridge
  University Press)

\bibitem{young1998}
Young A~P (ed) 1998 {\em Spin glasses and random fields\/} (Singapore: World
  Scientific)

\bibitem{sherrington1975}
Sherrington D and Kirkpatrick S 1975 {\em Phys. Rev. Lett.\/} {\bf 35} 1792

\bibitem{parisi1979}
Parisi G 1979 {\em Phys. Rev. Lett.\/} {\bf 43} 1754

\bibitem{parisi1983}
Parisi G 1983 {\em Phys. Rev. Lett.\/} {\bf 50} 1946

\bibitem{rammal1986}
Rammal R, Toulouse G and Virasoro M~A 1986 {\em Rev. Mod. Phys.\/} {\bf 58} 765

\bibitem{young1983}
Young A~P 1983 {\em Phys. Rev. Lett.\/} {\bf 51} 13

\bibitem{parisi1993}
Parisi G, Ritort F and Slanina F 1993 {\em J. Phys. A\/} {\bf 26} 3775

\bibitem{billoire2003}
Billoire A, Franz S and Marinari E 2003 {\em J. Phys. A\/} {\bf 36} 15

\bibitem{sinova2000}
Sinova J, Canright G and MacDonald A~H 2000 {\em Phys. Rev. Lett.\/} {\bf 85}
  2609

\bibitem{sinova2001}
Sinova J, Canright G, Castillo H and MacDonald A~H 2001 {\em Phys. Rev. B\/}
  {\bf 63} 104427

\bibitem{hed2004}
Hed G, Young A~P and Domany E 2004 {\em Phys. Rev. Lett.\/} {\bf 92} 157201

\bibitem{biroli2000}
Biroli G, Monasson R and Weigt M 2000 {\em Eur. Phys. J. B\/} {\bf 14} 551

\bibitem{mezard2002}
M\'ezard M, Parisi G and Zecchina R 2002 {\em Science\/} {\bf 297} 812

\bibitem{mezard2002b}
M\'ezard M and Zecchina R 2002 {\em Phys. Rev. E\/} {\bf 66} 056126

\bibitem{mertens1998}
Mertens S 1998 {\em Phys. Rev. Lett.\/} {\bf 81} 4281

\bibitem{mertens2000}
Mertens S 2000 {\em Phys. Rev. Lett.\/} {\bf 84} 1347

\bibitem{mulet2002}
Mulet R, Pagnani A, Weigt M and Zecchina R 2002 {\em Phys. Rev. Lett.\/} {\bf
  89} 268701

\bibitem{cover2000}
Weigt M and Hartmann A~K 2000 {\em Phys. Rev. Lett.\/} {\bf 84} 6118

\bibitem{cover-tcs2001}
Hartmann A~K and Weigt M 2001 {\em Theor. Comp. Sci.\/} {\bf 265} 199

\bibitem{cover-long2001}
Weigt M and Hartmann A~K 2001 {\em Phys. Rev. E\/} {\bf 63} 056127

\bibitem{zhou2003}
Zhou H 2003 {\em Eur. Phys. J. B\/} {\bf 32} 265

\bibitem{cover-review2003}
Hartmann A~K and Weigt M 2003 {\em J. Phys. A\/} {\bf 36} 11069

\bibitem{monasson1996}
Monasson R and Zecchina R 1996 {\em Phys. Rev. Lett.\/} {\bf 76} 3881

\bibitem{monasson1999}
Monasson R, Zecchina R, Kirkpatrick S, Selman B and Troyansky L 1999 {\em
  Nature\/} {\bf 400} 133

\bibitem{weigt2004}
Weigt M 2004 {\em New Optimization Algorithms in Physics\/} ed Hartmann A~K and
  Rieger H (Weinheim: Wiley-VCH)

\bibitem{zecchina2004}
Zecchina R 2004 {\em New Optimization Algorithms in Physics\/} ed Hartmann A~K
  and Rieger H (Weinheim: Wiley-VCH)

\bibitem{mezard2003}
M\'ezard M, Ricci-Tersenghi F and Zecchina R 2003 {\em J. Stat. Phys.\/} {\bf
  111} 505

\bibitem{bauer2001}
Bauer M and Golinelli O 2001 {\em Eur. Phys. J. B\/} {\bf 24} 339

\bibitem{bauer2001a}
Bauer M and Golinelli O 2001 {\em Phys. Rev. Lett.\/} {\bf 86} 2621

\bibitem{vccluster2004}
Barthel W and Hartmann A~K 2004 {\em Phys. Rev. E\/} {\bf 70} 066120

\bibitem{opt-phys2001}
Hartmann A~K and Rieger H 2001 {\em Optimization Algorithms in Physics\/}
  (Weinheim: Wiley-VCH)

\bibitem{gent1996}
Gent I~P and Walsh T 1996 {\em Proceedings of 12th European Conference on
  Artificial Intelligence. ECAI '96\/} (Chichester: Wiley) p 170

\bibitem{borgs2001}
Borgs C, Chayes J and Pittel B 2001 {\em Rand. Struct. Alg.\/} {\bf 19} 247

\bibitem{korf1998}
Korf R~E 1998 {\em Artif. Intell.\/} {\bf 106} 181

\bibitem{lawler66}
Lawler E~L and Wood D~E 1966 {\em Oper. Res.\/} {\bf 14} 699

\bibitem{phase-transitions2005}
Hartmann A~K and Weigt M 2005 {\em Phase Transitions in Combinatorial
  Optimization Problems\/} (Weinheim: Wiley-VCH)

\bibitem{tarjan1977}
Tarjan R~E and Trojanowski A~E 1977 {\em SIAM J. Comp.\/} {\bf 6} 537

\bibitem{mertens2006}
Mertens S 2006 {\em Computational Complexity and Statistical Physics\/} ed
  Percus A~G, Istrate G and Moore C (New York: Oxford University Press) p 125

\bibitem{valleys-long}
Hartmann A~K 2001 {\em Phys. Rev. E\/} {\bf 63} 016106

\bibitem{jain1988}
Jain A~K and Dubes R~C 1988 {\em Algorithms for Clustering Data\/} (Englewood
  Cliffs, USA: Prentice-Hall)

\bibitem{hed2001}
Hed G, Hartmann A~K, Stauffer D and Domany E 2001 {\em Phys. Rev. Lett.\/} {\bf
  86} 3148

\bibitem{ciliberti2003}
Ciliberti S and Marinari E 2004 {\em J. Stat. Phys.\/} {\bf 115} 557

\bibitem{ward1963}
Ward J 1963 {\em J. of the Am. Stat. Association\/} {\bf 58} 236

\bibitem{bauke2004}
Bauke H, Franz S and Mertens S 2004 {\em JSTAT\/} {\bf 4} P04003

\bibitem{derrida1981}
Derrida B 1991 {\em Phys. Rev. B\/} {\bf 24} 2613

\end{thebibliography}

\end{document}